\documentclass[10pt,conference]{IEEEtran}
\IEEEoverridecommandlockouts
\usepackage[utf8]{inputenc}
\usepackage{cite}
\usepackage{amsmath,amssymb,amsfonts}
\usepackage{algorithm,algorithmic}
\usepackage{graphicx}
\usepackage{textcomp}
\usepackage{xcolor}
\usepackage{siunitx}
\usepackage{booktabs}
\usepackage{url}
\usepackage{subcaption}
\usepackage{stfloats}
\usepackage[capitalize]{cleveref}
\usepackage{multirow}
\usepackage{upgreek}
\usepackage{mathtools}
\usepackage[left=1.71cm,right=1.71cm,top=2cm,bottom=2.6cm]{geometry}
\mathtoolsset{centercolon}

\usepackage{pgfplots}
\pgfplotsset{compat=newest}
\newlength\fwidth

\pgfdeclarelayer{background}
\pgfsetlayers{background,main}
\usepgfplotslibrary{groupplots}

\definecolor{plotclblue}{RGB}{57,106,180}
\colorlet{plotclblue}{black!20!plotclblue}
\definecolor{plotclorange}{RGB}{218,134,48}
\colorlet{plotclorange}{black!5!plotclorange}
\definecolor{plotclgreen}{RGB}{62,150,81}
\colorlet{plotclgreen}{black!10!plotclgreen}
\definecolor{plotclred}{RGB}{204,37,41}
\colorlet{plotclred}{black!10!plotclred}
\definecolor{plotclblack}{RGB}{83,81,84}
\colorlet{plotclblack}{black!10!plotclblack}
\definecolor{plotclviolet}{RGB}{107,76,154}
\colorlet{plotclviolet}{black!10!plotclviolet}
\definecolor{plotclredbrown}{RGB}{146,36,40}
\colorlet{plotclredbrown}{black!10!plotclredbrown}
\definecolor{plotclocer}{RGB}{148,139,61}
\colorlet{plotclocer}{black!10!plotclocer}
\definecolor{plotclyellow}{RGB}{255,255,30}
\colorlet{plotclyellow}{black!20!plotclyellow}
\definecolor{plotclcyan}{rgb}{000000,0.70000,0.70000}
\colorlet{plotclcyan}{black!20!plotclcyan}

\usepackage{tikz}
\usetikzlibrary{calc}
\usetikzlibrary{decorations.pathmorphing}
\usetikzlibrary{decorations.markings}
\usetikzlibrary{decorations.pathreplacing}
\usetikzlibrary{positioning}
\usetikzlibrary{fit}
\usetikzlibrary{arrows}
\usetikzlibrary{patterns}
\usepackage{circuitikz}
\usetikzlibrary{shapes.geometric}

\makeatletter
\tikzset{every picture/.style={font issue=\footnotesize},
        font issue/.style={execute at begin picture={#1\selectfont}}
        }
\tikzset{
  nomorepostactions/.code={\let\tikz@postactions=\pgfutil@empty},
  mymark/.style 2 args={decoration={markings,
    mark= between positions 0 and 1 step (1/8)*\pgfdecoratedpathlength with{%
        \tikzset{#2,scale=1,every mark}\tikz@options
        \pgftransformresetnontranslations
        \pgfuseplotmark{#1}%
      },  
    },
    postaction={decorate},
    /pgfplots/legend image post style={
        mark=#1,mark options={#2, scale=1},every path/.append style={nomorepostactions}
    },
  },
}

\tikzset{%
    block-common/.style={draw, semithick, fill=white, minimum height=2em, minimum width=2em},
    block/.style={rectangle, block-common},
    txtblock/.style={block, align=center, minimum height=4em},
    txtblocktight/.style={block, align=center, minimum height=2.3em},
    input/.style={inner sep=1pt},
    output/.style={inner sep=1pt},
    sum/.style = {draw, fill=white, circle, minimum size=1.1em, inner sep=0pt,
      font={\small$+$}},
    prod/.style = {draw, fill=white, circle, minimum size=1.1em, inner sep=0pt,
      font={\normalsize$\times$}},
    pinstyle/.style = {pin edge={to-,thin,black}}
}

\newcommand{\p}{\prime}

\usepackage[draft,nomargin,marginclue,inline]{fixme}

\renewcommand*\FXLayoutMarginClue[3]{%
  \marginpar[%
  \raggedleft\@fxuseface{margin}\textcolor{black}{\ignorespaces#3 fixme}]{%
    \raggedright\@fxuseface{margin}\textcolor{black}{\ignorespaces#3 fixme}}}
\makeatother

\begin{document}

\title{Exploiting Beamforming for Enforcing Semantic Secrecy in 5G NR mmWave Communications
\thanks{This work was supported by the German Ministry of Education and
Research (BMBF) within the national initiative for Post-Shannon
Communication NewCom (Grant 16KIS1003K). H. Boche, U. M\"onich and L. Torres-Figueroa were further supported in part by the BMBF as part of the \mbox{6G-life} project (Grant 16KISK002). The authors also received funding from the
Bavarian Ministry of Economic Affairs, Regional Development and Energy as part of the
 the 6G Future Lab Bavaria project.}%
}

\author{%
    \IEEEauthorblockN{%
    Luis Torres-Figueroa\IEEEauthorrefmark{1}, 
    Johannes Voichtleitner\IEEEauthorrefmark{1},  
    Ullrich J. M{\"o}nich\IEEEauthorrefmark{1},
    Taro Eichler\IEEEauthorrefmark{2},\\
    Moritz Wiese\IEEEauthorrefmark{1}, 
    Holger Boche\IEEEauthorrefmark{1}    
    }
    \IEEEauthorblockA{
    \IEEEauthorrefmark{1} 
     \textit{Chair of Theoretical Information Technology, Technical University of Munich}, Munich, Germany,\\
    \IEEEauthorrefmark{2} 
    \textit{Rohde \& Schwarz GmbH \& Co. KG}, Munich, Germany\\
    \{luis.torres.figueroa, johannes.voichtleitner, moenich, wiese, boche\}@tum.de, 
    taro.eichler@rohde-schwarz.com}%
}

\maketitle

\begin{abstract}

We experimentally investigate 
the performance of semantically-secure 
physical layer security (PLS) in 5G new radio (NR) mmWave communications during the initial cell search procedure in the NR band n257 at 27 GHz. A gNB transmits PLS-encoded messages 
in the presence of an eavesdropper, who intercepts the communication by non-intrusively collecting channel readings in the form of IQ samples. For the message transmission, we use the physical broadcast channel (PBCH) within the synchronization signal block. We analyze different signal-to-noise ratio (SNR) conditions by progressively reducing the transmit power of the subcarriers carrying the PBCH channel, while ensuring optimal conditions for over-the-air frequency and timing synchronization.  
We measure the secrecy performance of the communication in terms of upper and lower bounds for the distinguishing error rate (DER) metric for different SNR levels and beam angles when performing beamsteering in indoor scenarios, such as office environments and laboratory settings. 

\end{abstract}

\begin{IEEEkeywords}
Physical layer security,  wiretap channel, 5G NR,  mmWave communications, semantic security, beamforming.
\end{IEEEkeywords}

\section{Introduction}

Building upon the new capabilities of 5G new radio (NR) specifications, the next 6th generation (6G) of cellular communications aims to support an increased traffic demand coming from diverse verticals with more stringent requirements, such as in industrial or safety-critical communications, which makes the radio access network more susceptible to malicious attacks. 

5G NR copes with higher data rates by supporting wireless links at higher carrier frequencies in the millimeter wave (mmWave) region, known as frequency range 2 (FR2), where larger bandwidths up to \SI{2}{GHz} are currently specified \cite[Section 5.3]{3gpp.38.1012}. 
In FR2 and at higher frequencies, 
however, multi-antenna beamforming capabilities are needed to counter the increased path loss.
For this, 5G NR adopts \textit{analog beamforming} in FR2  
due to its reduced complexity and lower cost compared to \textit{digital beamfoming}, which is preferred in sub-\SI{6}{GHz} FR1 bands 
 \cite{dreifuerstMassiveMIMO5G2023}. 

While 5G NR specifies procedures for beam management and recovery to ensure reliability in the communication, the confidentiality still relies on cryptographic algorithms, some of which are regarded as compromised in a post-quantum era. 
Since trustworthiness represents a keystone in the development of 6G \cite{fettweis20226g}, it is envisioned that 6G will support an extra layer of security by exploiting channel characteristics to embed physical layer security (PLS) in the protocol stack, which offers information-theoretical guarantees \cite{schaefer2017information}. 

Thanks to its inherent beam-based approach, mmWave communications exhibit better PLS secrecy performance  compared to its sub-\SI{6}{GHz} counterparts \cite{wangSecureCommunicationCellular2014}. However, this has been mainly investigated analytically in terms of the ergodic secrecy rate \cite{wangSecureCommunicationCellular2014,wang_pls_in_mmwave_2016}, i.e., the maximum achievable secrecy rate based on channel state information (CSI), as measured by the signal-to-noise ratio (SNR), and lacking explicit code constructions. 

With the introduction of \textit{semantic security} in \cite{bellareCryptographicTreatmentWiretap2012}, a more rigorous security metric which can be quantified in practical settings thanks to its operational meaning has been used for PLS code design, whose performance in FR1  has been evaluated for  additive white Gaussian noise (AWGN) \cite{torres2021experimental,frank2022implementation,voichtleitner_statistical_2023} and fading channels \cite{10437737} by using a seeded modular coding scheme, this at the cost of a negligible latency overhead \cite{torres-figueroaImplementationPhysicalLayer2022}. 
Further, in \cite{9562229} semantic security is analytically evaluated for indoor terahertz (THz) scenarios based on the channel resolvability; although the authors do not use any explicit code construction.

In addition to increased privacy and security in the communication, resilience against jamming attacks is another central topic in 6G \cite{boche_mno_2023}. However, due to the complexity in the information-theoretic analysis of wiretap channels involving jamming attacks, it is still an open research question whether a modular coding scheme would additionally be effective against jamming  \cite{boche_jamming_2013,schaefer_jamming_2015}. Further, exploiting beamforming capabilities in mmWave and higher frequencies introduces also here interesting effects that need to be yet further investigated.

The present work aims to fill some of the gaps for supporting the integration of PLS in beam-based mmWave 6G systems. For this purpose, we investigate the secrecy performance in terms of upper and lower bounds on the distinguishing error rate (DER) when transmitting PLS-encoded messages using the physical broadcast channel (PBCH) during the 5G NR cell search procedure. For this, we use semantically-secure code constructions based on a seeded modular coding scheme that allows a smooth integration with existing transmission codes.  

By taking into account the effect of different antenna  directivity gains, reflected signals, beam misalignment, and dynamic side-lobe power levels due to beamsteering on the secrecy analysis of mmWave communications, to the best of our knowledge, this is the first work that experimentally investigates semantically-secure PLS in scattering rich environments 
considering real-life beam alignment and side-lobe issues caused when performing analog beamforming. 

The structure of this paper is as follows. 
Details on how semantically-secure PLS is implemented and measured are given in Section \ref{sec:it_pls}, 

while its integration into 5G NR's PBCH channel, mmWave deployment aspects, and investigated scenarios 
are presented in Section \ref{sec:integration_pls_in_5g}. Our experimental hardware setup is described in Section \ref{sec:hw_setup}, 
and the results 
 are discussed in Section \ref{sec:analysis}. Final conclusions are drawn in Section \ref{sec:conclusions}.

\section{Semantically-secure physical layer security} \label{sec:it_pls}

\subsection{Wiretap channel model}

Consider two legitimate parties, Alice and Bob, who communicate over a public insecure channel. 
We assume that an eavesdropper, Eve, is aware that a message exchange is taking place and it 
 intercepts the communication,  attempting to extract as much information as feasible out of its channel readings. In contrast to cryptographic approaches, our information-theoretic analysis poses no restriction on Eve's computational capabilities, which are deemed unlimited, adhering to a post-quantum era 
 scenario. 

Furthermore, we assume that Eve has full knowledge of the PLS codes, transmission codes, as well as of the low-level system parameters employed. 
The only two factors that Eve cannot alter to its benefit are 1) the channel conditions, and 2) the local randomness used by Alice at its PLS encoder. It is assumed that the channel conditions between Alice and Eve are worse than those between Alice and Bob. This difference in the channel conditions 
is exploited by the PLS encoder for embedding security in the communication.  

\begin{figure}[t]
  \centering
\resizebox{0.85\columnwidth}{!}{%
	{  \begin{tikzpicture}[font=\footnotesize, >=latex]
    \node[input] (M) {};
    \node[block, align=center, right=1.2em of M]
        (sec-layer-1)
        {\scriptsize $f_s^{-1}$};
    \node[block, align=center, right=1.5em of sec-layer-1]
        (ec-mod-layer-1)
        {\scriptsize Transm.\\ \scriptsize encoder};
    \node[block, right=2em of ec-mod-layer-1]
        (ch-bob)
        {\scriptsize $W_B$};
    \node[block, align=center, right=2.3em of ch-bob]
        (ec-mod-layer-2)
        {\scriptsize Transm.\\ \scriptsize decoder};
    \node[block, align=center, right=1.5em of ec-mod-layer-2]
        (sec-layer-2)
        {\scriptsize $f_s$}; 

    \node[output, right=1.2em of sec-layer-2] (M-hat) {};

    \node[block, below right=1.7em and 2em of ec-mod-layer-1]
        (ch-eve)
        {\scriptsize $W_E$};
    \node[block, right=2.3em of ch-eve] (eve) {$\mathcal{A}$};

    \draw[->] (M)
        -- node[above, xshift=-0.5em]   {$m$} (sec-layer-1);
    \draw[->] (sec-layer-1)
        -- node[above]                  {$v$} (ec-mod-layer-1);
    \draw[->] (ec-mod-layer-1)
        -- coordinate (mid) node[above] {$x$} (ch-bob);
    \draw[->] (ch-bob)
        -- node[above, xshift=-0.3em]   {$y$} (ec-mod-layer-2);
    \draw[->] (ec-mod-layer-2)
        -- node[above]                  {$\hat{v}$} (sec-layer-2);
    \draw[->] (sec-layer-2)
        -- node[above, xshift=0.8em]    {$\hat{m}$} (M-hat);    
    \draw[->] (mid)
        |- (ch-eve);
    \draw[->] (ch-eve)
        -- node[above, xshift=0.1em]   {$z$} (eve);
    
    \node[draw, rectangle, densely dashed, inner xsep=0.4em, inner ysep=0.35em, yshift=-0.13em,
    fit=(sec-layer-1)(ec-mod-layer-1)] (alice) {};
    \node[draw, rectangle, densely dashed, inner xsep=0.4em,
    inner ysep=0.35em, yshift=-0.13em, xshift=-0.24em,
    fit=(sec-layer-2)(ec-mod-layer-2)] (bob) {};
    \node[draw, rectangle, densely dashed, inner ysep=0.35em, inner xsep=0.4em, yshift=-0.13em,
    fit=(eve)] (eve1) {};
    \node[anchor=south west, xshift=-1.5em] at (alice.north west) {Legitimate sender: Alice (gNB)};
    \node[anchor=south west, xshift=-1.5em] at   (bob.north west) {Legitimate receiver: Bob (UE)};
    \node[anchor=north west, xshift=-1.5em] at   (eve1.south west) {Eavesdropper: Eve};

    \node[below=1.5em of ch-eve, align=center] (seed) {Seed $s$};
    \draw[->] (seed) -| (sec-layer-1);
    \draw[->] (seed) -| (sec-layer-2);
    \path let \p1=(seed.east), \p2=(eve) in coordinate (split) at (\x2,\y1);
    \draw[->, densely dotted] (split) -- (eve);

  \end{tikzpicture}}
  }%
  \caption{Seeded modular coding scheme for the wiretap channel $(W_{B},W_{E})$, showing its component secrecy functions ($f_s^{-1}$,$f_s$) and transmission codes.}
  \label{fig:wiretap_ch}
  \vspace{-1em}
\end{figure}
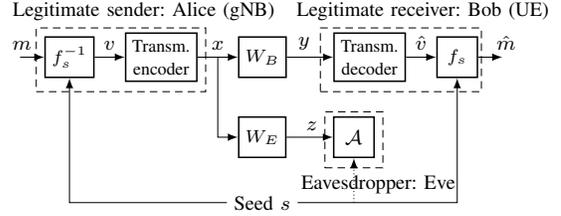

The mathematical model for analyzing such communication settings is the wiretap channel model introduced by Wyner	\cite{Wyner}, which has been generalized for discrete memoryless channels \cite{CK} and further extended to Gaussian channels \cite{LH}. 
\cref{fig:wiretap_ch} depicts the latter case, where the transmitter Alice sends a message $m$ to the legitimate receiver Bob via the Gaussian channel $W_{B}$, while an eavesdropper Eve intercepts the communication over a noisier Gaussian channel $W_{E}$. 

\subsection{Modular coding scheme}

In order to minimize the information leakage to Eve, Alice and Bob use 
a modular coding scheme \cite{Bellare2012a}, which allows us to 
design secrecy codes 
separately from the transmission codes. The main advantage of the modular coding scheme is that the secrecy rate can be dynamically adapted depending on the radio conditions of Eve. This is possible thanks to the use of a seed information vector $s$, which is transmitted unprotected over the channel, i.e., Eve may intercept and use it as part of its attack strategy. Similarly, the transmission rate can be separately adapted depending on the radio conditions of Bob. 

This facilitates a flexible deployment of a variety of secrecy codes, independent of the transmission code being used.

\subsection{Explicit PLS code construction} \label{sec:sec_func}

We employ the code construction introduced in \cite{hayashi_exponential_2011}, which outperforms other existing secrecy encoders  \cite{voichtleitner_gc23} and achieves \emph{semantic security} \cite{bellareCryptographicTreatmentWiretap2012}. This is, to the best of our knowledge, the strongest secrecy metric in the literature, since it does not pose any restriction on Eve's attack strategies. 

The secrecy decoding function at the receiver Bob is constructed using a universal hash function $f_{s}:\{0,1\}^l \rightarrow \{0,1\}^k$ with seed $s = (a,t) \in \mathbb{F}_2^{l-1} \times \mathbb{F}_2^{k}$, defined as
\begin{equation}
  f_{s}(\hat{v})
  =
 [A_{\textit{a}},I]\hat{v}+t, 
\end{equation}
where $A_{\textit{a}}$ is a random Toeplitz matrix of size $k \times (l-k)$ generated with a random seed $\textit{a}$ of length $(l-1)$ bits, $I$ is the identity matrix of dimension $k$, and $\hat{v} \in \{0,1\}^l$ is the output of the channel decoder.

At the transmitter, 
the encoding function is the stochastic inverse function $f_{s}^{-1}: \{0,1\}^{l-k} \times \{0,1\}^k \rightarrow \{0,1\}^l$, given as
\begin{equation}
  f_{s}^{-1}(R,m)
  =
 [R,(A_{\textit{a}}R)+m+t]^{\intercal},
\end{equation}
where $R \in \{0,1\}^{l-k}$ is an $(l-k)$ binary random variable with a uniform distribution and $m \in \{0,1\}^k$ is the message of length $k$ \SI{}{bits} to be transmitted.

\subsection{Secrecy Metrics} \label{sec:secrecy_metrics}

\subsubsection{Distinguishing error rate}

Distinguishing security (DS), as defined in \cite{Bellare2012a}, represents a form of \emph{semantic security} \cite{bellareCryptographicTreatmentWiretap2012}. While the traditional definition of semantic security considers the probability distribution of the entire set of all messages $\mathcal{M}$, 
DS considers for the sake of the secrecy analysis all possible message pairs instead, 
without loss of generality, thus simplifying the inherent complexity in the secrecy analysis. 
In this paper, we use an equivalent security metric: The 
DER \cite{frank2022implementation}. 
When considering the message pair ($m_1$, $m_2$) together with the seed $s$, the DER can be then expressed as 
\begin{equation}
  \text{DER}_\text{E}(s,m_1,m_2)
  =
 \min_{\mathcal{A}} \Pr[\mathcal{A}(s,m_1, m_2, z(m_\Theta)) \!\neq\! \Theta],
\end{equation}
where $\Theta$ is a uniformly distributed random variable over $\{1,2\}$, $z(m_{\Theta})$ is the channel output at the eavesdropper when message $m_{\Theta}$ is sent, and $\mathcal{A}$ the attack strategy of the eavesdropper. The choice of one single message pair for the secrecy analysis is further discussed in \cref{sec:msg_lengh_discussion}.

Under this paradigm, the communication system achieves 
a higher security level the closer the DER approaches $0.5$. Conversely, 
the security decreases as the DER approaches $0$.

The optimal attack strategy $\mathcal{A}$ at Eve for AWGN channels is the maximum likelihood decoder, whose complexity grows exponentially with the encoding randomness length $(l-k)$, reaching prohibitively long computation times for the parameters in \cref{tab:pls_params} used in our experiments. Therefore, we do not employ a maximum likelihood decoder, but an upper and lower bound on the maximum likelihood decoder instead, as discussed in the following subsection. 

\subsubsection{Upper and lower bounds for the DER based on SCL} \label{sec:up_low_bound}

Given the complexity of the DER computation, upper and lower bounds were introduced in \cite{voichtleitner_statistical_2023}. In this paper, we use a method which employs a successive cancellation list (SCL) decoder  \cite{tal2015list} as a list generator to evaluate the secrecy performance in our experiments. First, we use the SCL decoder to generate a list. 
We reduce the list by all codewords that cannot be assigned to the messages $m_1$ or $m_2$. The upper bound decides on the most likely codeword from the remaining codewords for the given channel output. The lower bound compares the actual transmitted codeword with the decision of the upper bound and takes the most likely of these two. As the list size $L$ increases, the upper and lower bounds come  
 closer together, and the verification of the secrecy assurance is more precise, at the cost of increasing computation times. 

\section{Integration of PLS in 5G NR FR2} \label{sec:integration_pls_in_5g}

\subsection{5G NR cell search procedure} \label{sec:beam_search_proc}

We focus our analysis on the initial downlink cell search procedure, when a base station gNB (Alice) sends synchronization signal blocks (SSBs) using multiple transmit beams.

\begin{figure}[t]
    \centering
    \includegraphics[width=.99\linewidth]{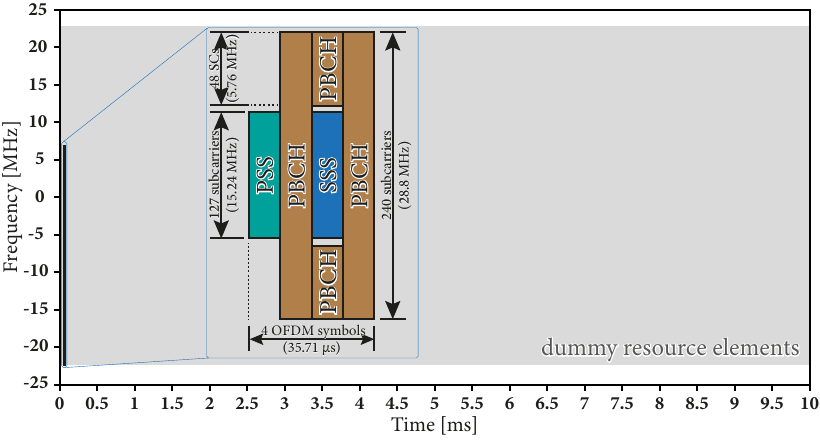}
    \caption{5G NR time-frequency frame structure used for secure message transmission via PBCH within the SSB\#0 (zoomed view). The gray region represents random (dummy) data sent with a relative transmit power of \SI{0}{dB}.}
    \label{fig:ssb_tf_grid}
\end{figure}

\begin{figure}[t]
\centering
\begin{minipage}{0.34\linewidth}
  \begin{subfigure}{\linewidth}
  \centering
  \includegraphics[width=\linewidth]{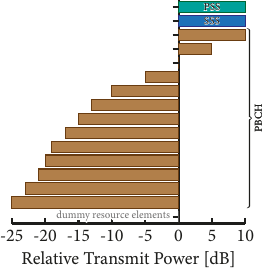}%
  \label{fig:relative_power_levels_v2}
  \end{subfigure}%
\end{minipage}
\begin{minipage}{0.64\linewidth}
  \begin{subfigure}{.9\linewidth}
  \centering
  \includegraphics[width=\linewidth]{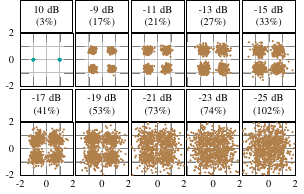}
  \label{fig:aces_const_diag_theta_0}
  \end{subfigure}%
  \vspace*{-1.4em}
\end{minipage}
\caption{Relative transmit power w.r.t. the absolute transmit power, $P_{\text{Tx}}$, in \cref{tab:scenarios} used for the SSB in \cref{fig:ssb_tf_grid} (left). The received PSS and PBCH constellation diagrams, for different $P_{\text{PBCH}}$ values in \SI{}{dB} for scenario 2 with $\theta=0$° are shown along with their error vector magnitude (EVM) in \% (right).}
\label{fig:rel_pwr_levels}
\end{figure}

The gNB transmits an m-sequence and a Gold sequence carried using the primary (PSS) and secondary synchronization signals (SSS) in the SSB block in \cref{fig:ssb_tf_grid}, respectively,  with a power level \SI{10}{dB} higher than the 
unused radio resources carrying dummy resource elements 
 in order to ensure optimal over-the-air downlink  timing and frequency synchronization for different SNR regimes. 
Conversely, the resource elements (REs), i.e., subcarriers lasting one OFDM symbol, that carry the 
PBCH and its associated demodulation reference signals (DM-RS) are transmitted using increasingly lower transmit power levels, as depicted in \cref{fig:rel_pwr_levels}, where also selected 
received constellation diagrams at Eve are shown. The actual relative power values used are listed in \cref{tab:hf_parameters}. 
This facilitates the analysis of a wide range of SNR levels, 
while ensuring a correct synchronization and RE demapping.

After the cell selection, the UE extracts the PBCH in the SSB, which is used for broadcasting system parameters via a master information block (MIB) with a fixed transmission rate $R_{\text{MIB}}=64/864$, using polar codes and QPSK modulation. 
The DM-RS multiplexed in the time and frequency domains with the PBCH is  
used for channel estimation and equalization. 
In our experiments, instead of carrying the MIB containing system parameters via the PBCH, we transmit randomly generated messages $m$ that are PLS-encoded using the secrecy code described 
in \cref{sec:sec_func}, before being appended with an \SI{11}{bit} CRC, and mapped to codewords of length $n$ using polar codes and a transmission rate $R=l/n$.

\begin{figure}[b]
    \centering
    \includegraphics[width=.99  \linewidth]{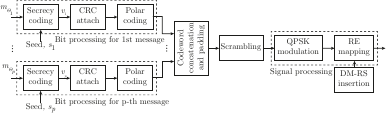}
    \caption{Modified physical-layer procedure for the PBCH at the transmitter, including a secrecy encoder and the concatenation of multiple codewords.}
    \label{fig:pbch_procedure}
\end{figure}

We append multiple codewords and add padding bits to fill the available \SI{864}{bits} that are carried by the PBCH, before doing scrambling, modulation, and RE mapping  in \cref{fig:pbch_procedure}.

\begin{figure*}[t!]
\centering
\begin{minipage}{0.31\linewidth}
  \begin{subfigure}{\linewidth}
  \centering
  \includegraphics[width=\linewidth]{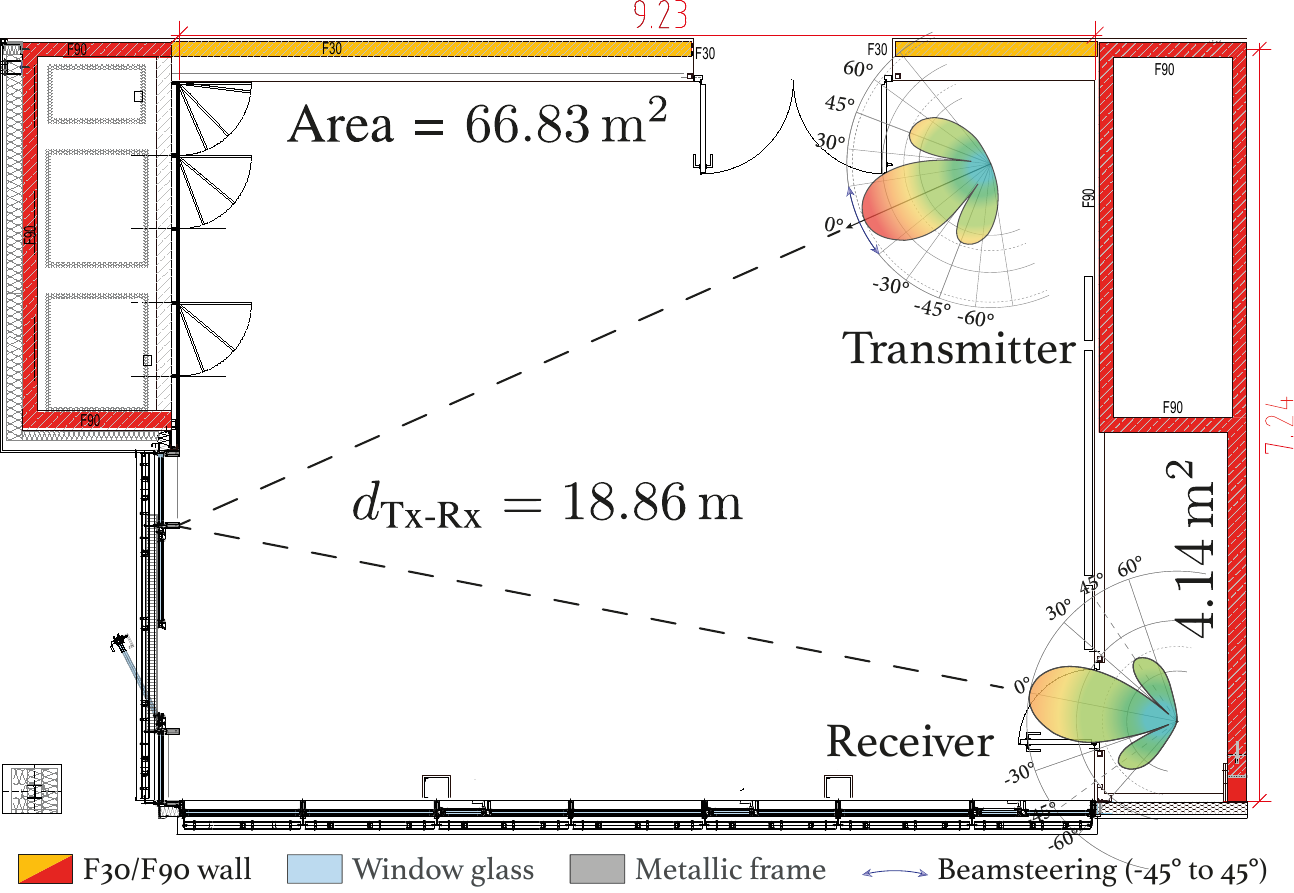}
  \subcaption{Scenario 1 (NLOS): Conference room at R\&S.}
  \label{fig:sc1_conf_room}
  \end{subfigure}%
\end{minipage}
\begin{minipage}{0.31\linewidth}
  \begin{subfigure}{\linewidth}
  \centering
  \includegraphics[width=0.87\linewidth]{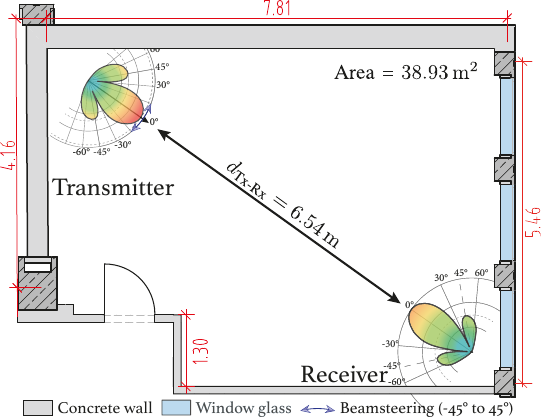}
  \subcaption{Scenario 2: Laboratory setting at TUM.}
  \label{fig:sc2_aces_lab}
  \end{subfigure}%
\end{minipage}
\begin{minipage}{0.34\linewidth}
  \begin{subfigure}{\linewidth}
  \centering
  \includegraphics[width=0.87\linewidth]{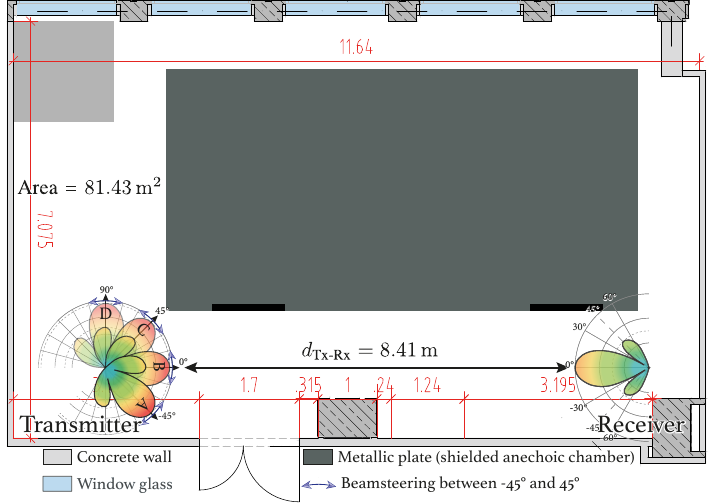}
  \subcaption{Scenario 3: Outside an anechoic chamber at TUM.}
  \label{fig:sc3_messraum}
  \end{subfigure}%
\end{minipage}
\caption{Blueprint of the investigated indoor NLOS and LOS indoor scenarios, showing the radiation pattern of the transmit and receive antennas.}
\label{fig:scenarios_desc}
\vspace*{-1.5em}
\end{figure*}

We analyze three indoor scenarios, as listed in \cref{tab:scenarios} and depicted in \cref{fig:scenarios_desc}, where the receiver, acting as an eavesdropper, has privileged access to the location 
where the communication takes place. In each scenario, the transmitting gNB performs beam search by means of analog beamsteering between -45° and +45°. For scenario 3, 
we additionally evaluate different azimuth 
rotations at Alice: $0$°,$\pm45$°, and $90$°.

For eavesdroppers located outside, preliminary measurements showed that our 5G NR FR2 transmissions in the aforementioned scenarios 
were completely absorbed by concrete walls, wooden doors, and glass windows, to the extend that a signal 
detection via PSS outside of the rooms was not possible.

For the cell search procedure, we perform burst transmissions of the SSB 
in \cref{fig:ssb_tf_grid} with a  periodicity of \SI{10}{ms} using an OFDM waveform that occupies a bandwidth of \SI{50}{MHz}, and with a subcarrier spacing of \SI{120}{kHz}. 
We concatenate three different codewords of length $n=256$ bits each and add $96$ padding bits, before  scrambling, modulating, and mapping them onto the PBCH, as previously discussed.

We send the SSB $N_{\text{Tx}}$  times per relative PBCH power level, $P_{\text{PBCH}}$, and beam angle, $\theta$, summing up to \SI{93}{k}, \SI{61}{k}, and \SI{114}{k} transmissions for scenarios 1, 2 and 3 in \cref{tab:scenarios}, respectively.

For comparison 
between transmissions at different SNR levels, we randomly generate $3N_{\text{Tx}}$ codewords that are sequentially transmitted in triplets for each $P_{\text{PBCH}}$ and beam angle $\theta$. Since we store them, we have full knowledge of the PBCH's IQ symbols used at each burst transmission, which are employed at the receiver for estimating the SNR of the channel actually experienced by each individual RE in the PBCH.

\subsection{Message length for assessing upper and lower DER bounds} \label{sec:msg_lengh_discussion}
In all scenarios, we transmit a single bit per codeword, i.e., the message length $k$ is \SI{1}{bit}. 
The reason for this is to  
reduce the complexity when estimating the upper and lower bounds for the DER. 
By choosing $k=1$, we ensure that all codewords in the list belong to either $m_1$ or $m_2$. This reduces the size of the list 
needed for assessing the security level, as well as 
the computation time. In \cite{voichtleitner_statistical_2023} it was shown that the DER performance of Eve for fixed codeword length $n$ and fixed modulation order $M$ depends only on $(l-k)$. That is, 
the security analysis herein is valid for $n=256$, QPSK modulation and any message length $k>1$ and $l-k=221$. Note that the upper and the lower bounds are always $\le 0.5$.  

\subsection{Estimation of the signal-to-noise ratio} 

A statistical analysis of our measurements in the indoor environments in \cref{fig:scenarios_desc} showed a Gaussian-like distribution for $\text{SNR}_i$ over time for any RE in the PBCH, while the noise varies among subcarriers (see \cref{sec:sc2_results}). Given this quasi-stationary behavior,  we use in \cref{sec:analysis} 
the time-averaged SNR per RE, 
$\text{SNR}_i^{\text{avg}}$, given as
\begin{equation} \label{eq:snr_per_sc}
  \text{SNR}_i^{\text{avg}} = \frac{\bar{P}_i}{\bar{\upvarsigma}_{i}^2}
  =
 \frac{1}{N} \sum_{j=1}^{N} \frac{P_{i,j}}{\upvarsigma_{i,j}^2}, 
\end{equation}
where, for a given $P_{\text{PBCH}}$ and beam angle $\theta$, $P_{i,j}$ is the signal power of the $i$-th subcarrier within the $j$-th transmitted OFDM symbol, $\upvarsigma_{i,j}^2$ is the noise variance associated to it,  
 and $N$ is the total number of transmitted OFDM symbols.

Also, in our measurements we observe that the channel remains time-invariant 
during the $\SI{26.79}{\mu s}$ that the transmission of the three OFDM symbols carrying the PBCH channel lasts. Thus, we consider only OFDM symbols 1 and 3 in \cref{fig:ssb_tf_grid}, 
disregarding OFDM symbol 2 since its radio resources are shared with the SSS signal. 
That is, for the analysis we consider two OFDM symbols per SSB transmission; therefore  $N$ is twice the number of signal transmissions, $N=2 N_{\text{Tx}}$.

\section{Experimental 5G NR mmWave setup} \label{sec:hw_setup}

\subsection{Hardware setup}

\begin{figure}[t]
  \centering
  \includegraphics[width=.98\linewidth]{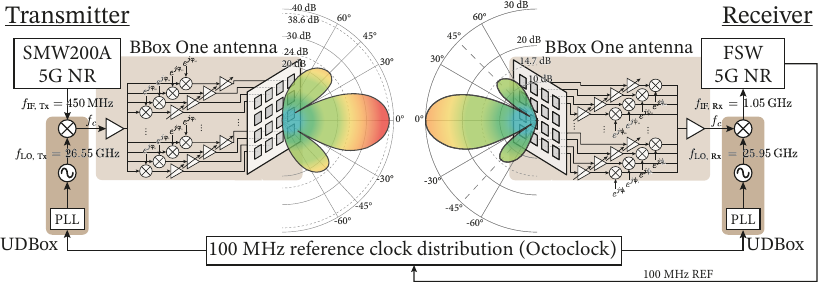}
  \caption{Experimental mmWave setup showing the measured radiation patterns including the main and side lobes of the transmit and receive BBox antennas, when the beam is steered in the boresight direction ($\theta=0$).}
  \vspace*{-1.5em}
  \label{fig:Experimental_setup}
\end{figure}

Our experimental setup is shown in \cref{fig:Experimental_setup}. At the transmitter, we employ an R\&S®SMW200A vector signal generator equipped with the SMW-K144 5G NR option. 
At the receiver, an R\&S®FSW26 signal and spectrum analyzer with the FSW-K144 5G NR option is used. 
For operating in FR2 at \SI{27}{GHz}, we use a  TMYTEK's mixer with internal local oscillator (UDBox) and a mmWave antenna module (BBox One 5G) at each communication end. We also use the FSW26 to generate a \SI{100}{MHz} reference signal  
with an accuracy of around \SI{0.1}{ppm}, 
that is distributed to the UDBoxes for clock synchronization via an NI CDA-2900 octoclock.

We developed a framework in C++ that uses SCPI commands for controlling the FSW26 and SMW200A via a wired management network, and TMYTEK's TLKCore API for configuring the UDBox and BBox parameters, including running the beamsteering algorithm and transmit power control. 

\subsection{Unwanted image frequency and analog beamsteering}

Due to its lack of internal bandpass filters, the mixing operation at the UDBox lets unfiltered image signals pass to the BBox One 5G. To mitigate them, we use asymmetric intermediate frequencies, $f_{\text{IF}}$, at the transmitter and receiver, as depicted in \cref{fig:Experimental_setup}. This prevents an overlap of the wanted signal and the downconverted image signal at the receiver, which is finally filtered out at baseband by the FSW26. 

At the transmitting gNB, we perform beamsteering by individually 
controlling the phase and gain of each antenna element of the BBox's $4\times4$ patch antenna array, with antenna spacing $\Delta d = \SI{5}{mm}$.
At the receiver (Eve), no beamsteering is done. Since the transmitter's position is fixed in all scenarios, we assume that the Eve knows it, and physically points its antenna in Alice's direction, with its beam directed to $\theta = 0$°. 

\subsection{Link budget for over-the-air transmissions}

\begin{figure}[t]
  \centering
\resizebox{0.95\columnwidth}{!}{%
	{\begin{tikzpicture}[font=\normalsize, >=latex]
\pgfdeclareimage[width=\linewidth]{img}{tikz/link_budget_devices}
\node (img1) at (0,0) {\pgfuseimage{img}};

\node[inner sep=0.2em] (smw) at (-3.7,-1) {\small$P_{\text{Tx}}$};
\node[inner sep=0.2em, right=2.2em of smw] (ud1) {\small$L_{\text{UD}}$};
\draw[->] (smw) -- node[above] {\small$L_{\text{IF,Tx}}$} (ud1);
\node[inner sep=0.2em, right=2.2em of ud1] (bb1) {\small$G_{\text{BB,Tx}}$};
\draw[->] (ud1) -- node[above] {\small$L_{\text{HF}}$} (bb1);
\node[inner sep=0.2em, right=4.0em of bb1] (bb2) {\small$G_{\text{BB,Rx}}$};
\draw[->] (bb1) -- node[above] {\small$L_{\text{PL}}$} (bb2);
\node[inner sep=0.2em, right=2.2em of bb2] (ud2) {\small$L_{\text{UD}}$};
\draw[->] (bb2) -- node[above] {\small$L_{\text{HF}}$} (ud2);
\node[inner sep=0.2em, right=2.2em of ud2] (fsw) {\small$P_{\text{Rx}}$};
\draw[->] (ud2) -- node[above] {\small$L_{\text{IF,Rx}}$} (fsw);
\end{tikzpicture}}
  }%
  \caption{FR2 link budget. The path loss (PL) depends on the beam alignment and geometry of each specific scenario. \cref{tab:scenarios} lists the boresight PL.}
  \label{fig:link_budget}
  \vspace{-1.5em}
\end{figure}
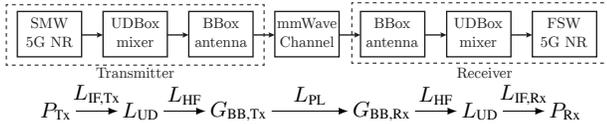

 To be able to cover large distances between transmitter and receiver, $d_{\text{Tx-Rx}}$, our setup operates at rather low IF frequencies and uses \SI{25}{m}-long low-loss CLF240 coaxial cables for connecting SMW200A and FSW26 to their corresponding UDBox, with a cable loss of $L_{\text{IF,Tx}} = \SI{4.33}{dB}$ 
 and $L_{\text{IF,Rx}} = \SI{6.68}{dB}$ 
 for all 
 scenarios. 
 The UDBox and BBox are connected using 
 a \SI{30.48}{cm}-long HP 160S coaxial cable with a loss of $L_{\text{HF}} \approx \SI{0.57}{dB}$. \cref{fig:link_budget} summarizes our link budget.

 The conversion loss introduced by the UDBox at both ends is $L_{\text{UD}} = \SI{13}{dB}$.  The corresponding overall antenna gains are set to $G_{\text{BB,Tx}} =\SI{51.68}{dB}$ and $G_{\text{BB,Rx}}=\SI{33.9}{dB}$, respectively.

 The free space path loss $L_{\text{PL}}$ in \cref{tab:scenarios} is calculated using  Friis's equation for the boresight beam ($\theta = 0$°) when the transmitter and receiver are aligned towards each other, 
 \begin{equation}
  L_{\text{PL}}
  =
 10 \log_{10} \left( \frac{4 \pi f_c d_{\text{Tx-Rx}} }{c} \right)^2 \quad \text{in \SI{}{dB}}, 
\end{equation}
where $f_c$ is the carrier frequency $d_{\text{Tx-Rx}}$ is the physical distance between transmit and receive antennas, 
and $c$ is the speed of light.

\begin{table}[b]
	  \vspace*{-1.4em}
\caption{Relative power levels used for transmission of SSB's components.}%
\centering
\begin{tabular}{@{}ccc@{}}
 \toprule
  Parameter & Variable & Value \\
 \midrule
Relative PBCH power & $P_{\text{PBCH}}$ & \{-25:2:+10\}\SI{}{dB} \\
Relative PSS/SSS power & \{$P_{\text{PSS}}$,$P_{\text{SSS}}$\} & \{10,10\}\SI{}{dB} \\
 \bottomrule
 	  \vspace*{-1.4em}
\end{tabular}
\label{tab:hf_parameters}
\end{table}

\begin{table}[b]
\caption{PLS parameters used in our experiments.}
\centering
\begin{tabular}{@{}lccc@{}}
 \toprule
  Parameter & Variable & Value & Unit\\
 \midrule
Message length & $k$ & $1$ & \SI{}{bit}\\
Seed component length & $l$ &  $222$ & \SI{}{bit} \\
Codeword length & $n$ & $256$ & \SI{}{bit}\\
Transmission rate & $R$ & $222/256 \approx 0.87$ & --\\ 
Secrecy rate & $k/(n/2)$ & $1/128 \approx \SI{0.008  }{}$ & \SI{}{bit/RE}\\ 
 \bottomrule
\end{tabular}
\label{tab:pls_params}
\end{table}

\begin{table}[b]
\vspace*{-1.4em}
\caption{Distance and link budget parameters for the evaluated scenarios.}%
\centering
\begin{tabular}{@{}clcccc@{}}
 \toprule
  ID & Description & $P_{\text{Tx}}$ & $d_{\text{Tx-Rx}}$ & $L_{\text{PL}}$ [\SI{}{dB}] & $N_{\text{Tx}}$\\
 \midrule
1 & Conference room & \SI{7}{dBm} & \SI{18.86}{m} &  \SI{-86.58}{} & 140\\
2 & Laboratory & \SI{-3}{dBm} & \SI{6.54}{m}  & \SI{-77.38}{} & 250 \\
3 & Measurement chamber & \SI{-3}{dBm} & \SI{8.41}{m} & \SI{-79.57}{} & 125\\
 \bottomrule
\end{tabular}
\label{tab:scenarios}
\end{table}

\section{Experimental results and analysis} \label{sec:analysis}

In each case,  \cref{fig:sc1_summary,fig:sc2_summary,fig:sc3_summary} show the upper and lower DER bounds for a SCL list size of $L=8$ (top) and the estimated SNR per beam angle, PBCH subcarrier, and $P_{\text{PBCH}}$ (bottom). 

\subsection{Scenario 1: Reflected signals in a conference room}

\begin{figure}[t!]
\centering
\begin{minipage}{\linewidth}
  \begin{subfigure}{\linewidth}
  \centering
	{\input{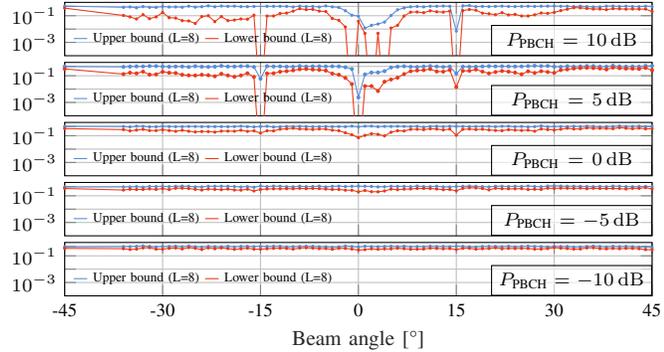}}
  \subcaption{Lower and upper DER bounds per beam angle and $P_{\text{PBCH}}$. The results for $P_{\text{PBCH}}=\{-15,-20,-25\}$ (not shown) are identical to $P_{\text{PBCH}}=-10$.}
  \label{fig:sc1_bounds}
  \end{subfigure}%
\end{minipage}
\begin{minipage}{\linewidth}
  \begin{subfigure}{0.99\linewidth}
  \centering
{\input{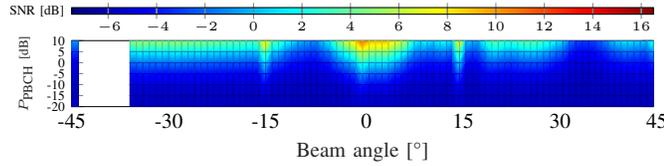}}
  \subcaption{SNR per relative power $P_{\text{PBCH}}$  and beam angle $\theta$.}
  \label{fig:info_leakage_sc1} 
  \end{subfigure}%
\end{minipage}
\caption{Measurements for scenario 1 (NLOS), where Eve exploits the reflected signals. No measurements were done in the blank region from -44° to -37°.}%
\label{fig:sc1_summary}
     \vspace*{-1.4em}
\end{figure}

In \cref{fig:sc1_conf_room}, Eve overhears Alice's transmissions via a non-line-of-sight (NLOS) channel by directing its beam in Alice's boresight beam direction, which lies on a metallic window frame. Eve keeps its beam fixed, while Alice performs beamsteering at steps $\Delta\theta=1$°. The measurements in \cref{fig:sc1_summary} show that the attenuated reflected signals under NLOS conditions for a virtual image distance of \SI{18.86}{m} still convey enough information when Alice's and Eve's beams are ``aligned'' and the absolute PBCH transmit power at the SMW200A lies between \SI{12}{} and \SI{17}{dBm}, accounting 
for a peak $\text{SNR}$ of  $\approx\SI{15}{dB}$. 
In such a case, Eve is able to exploit the reflected signals, as the DER bounds lie farther apart from $0.5$ in the two upper plots for $P_{\text{PBCH}}=\{10,5\}~\SI{}{dB}$. 
Even when the nulls at $\pm30$° are steered in Eve's direction, which reduces its 
SNR by $\approx \SI{10}{dB}$, this is not enough to entirely prevent information leakage, which suggests that enough reflected signals still arrive to Eve. 
Note that the different room materials (glass, F30/F90 walls, metallic frames) cause that the secrecy performance is not symmetric. 
Only as the alignment to the main and side lobes deviates further in the opposite direction from Eve's location, 
or the absolute PBCH transmit power lies below $\SI{7}{dBm}$, we observe higher secrecy assurance levels, as the upper and lower DER bounds  come closer together towards $0.5$.

\subsection{Scenario 2: Laboratory setting} \label{sec:sc2_results}

In \cref{fig:sc2_aces_lab}, Eve is now placed in the opposite corner in a laboratory room, with its antenna directed to Alice's. Given its closer proximity at \SI{6.54}{m}, line-of-sight (LOS) conditions, and despite a \SI{10}{dB} lower transmit power $P_{\text{Tx}}$ in \cref{tab:scenarios}, a weaker secrecy performance is measured as the transmitter steers its beam in Eve's direction for $P_{\text{PBCH}}>\SI{-23}{dB}$. The transmitter's side lobes do not play a significant role for an increased information leakage here, 
since the secrecy assurance increases monotonically as the beam misalignment increases. This effect may be due to the rich scattering environment in the laboratory. In this scenario, beam alignment to Eve needs to be avoided if transmitting PBCH power levels above \SI{-26}{dBm} if sufficient secrecy performance is to be guaranteed. Since we used a fixed set of PLS parameters, increasing the value $l-k$ should improve the secrecy performance, as shown in \cite{torres2021experimental}.

\begin{figure}[t!]
\centering
\begin{minipage}{\linewidth}
  \begin{subfigure}{\linewidth}
  \centering
	{\input{tikz/sc2_aces_der.tex}}
	  \vspace*{-1.4em}
  \subcaption{Lower and upper bounds on the DER per $\theta$ and $P_{\text{PBCH}}$. The security level weakens as $P_{\text{PBCH}} > -\SI{15}{dB}$ (not shown), except for $\theta=\pm30$° (nulls).}
  \label{fig:sc2_bounds}
  \end{subfigure}%
\end{minipage}
\begin{minipage}{\linewidth}
  \begin{subfigure}{\linewidth}
  \centering
  \includegraphics[width=\linewidth]{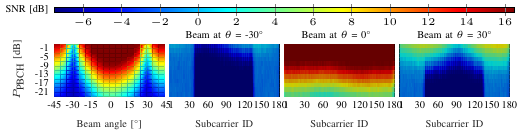}
	\vspace*{-1.4em}
  \subcaption{Estimated SNR in \SI{}{dB} per $P_{\text{PBCH}}$ and beam angle $\theta$ (leftmost). The frequency-selective behavior of the SNR per PBCH's subcarrier ID is shown on the right for three beam angles at $\pm30$° (nulls) and $0$° (main lobe).}
  \label{fig:snr_dist_sc2}
  \end{subfigure}%
\end{minipage}
\caption{Measurements for scenario 2 (laboratory setting), showing the lower and upper DER bounds  as well as the SNR per $\theta$ and $P_{\text{PBCH}}$. The frequency selectivity per subcarrier of the channel is shown for 3 selected beam angles.} 
\label{fig:sc2_summary}
      \vspace*{-1em}
\end{figure}

In contrast to the channel behavior in the time domain, 
our measurements in \cref{fig:sc2_summary} show a frequency-selective channel, where the SNR per subcarrier remains relatively constant over time, but fluctuates among different subcarriers. Further, as we decrease the $P_{\text{PBCH}}$, a gain drop is observed in the central frequency region where PSS and SSS are located at adjacent time-domain OFDM symbols, which may be acting as strong interferers, causing non-linear effects such as the desensitization of the weaker PBCH signal \cite[p.~32]{gu2006rf}.

\subsection{Scenario 3: Eavesdropper located next to the transmitter}
\begin{figure}[t!]
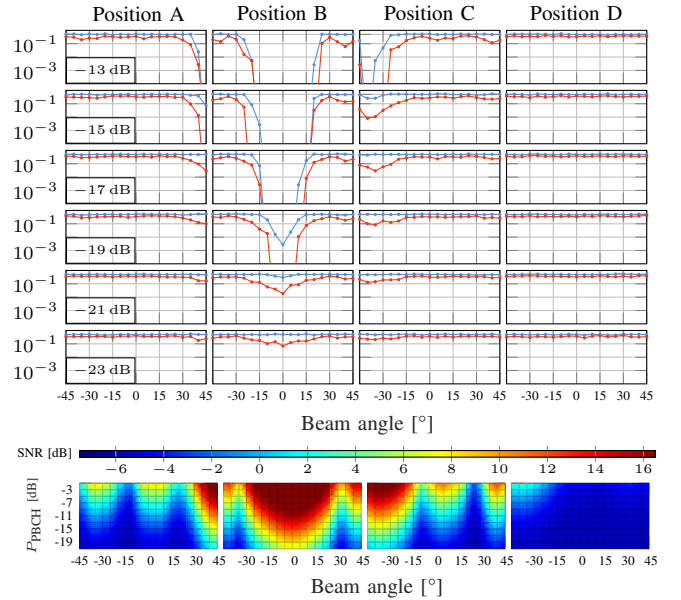

\centering
\begin{minipage}{\linewidth}
  \begin{subfigure}{\linewidth}
  \centering
      {\input{tikz/sc3_messraum_der.tex}}
  \label{fig:sc3_bounds}
  \end{subfigure}%
\end{minipage}
\begin{minipage}{0.98\linewidth}
  \begin{subfigure}{\linewidth}
  \centering
    {\input{tikz/sc3_messraum_pos_all_v2.tex}}
  \label{fig:snr_dist_sc3}
  \end{subfigure}%
      \vspace*{-1.6em}
\end{minipage}
\caption{Measurements for scenario 3 (outside an anechoic chamber), showing the aforementioned metric for the four positions depicted in \cref{fig:sc3_messraum}.} 
\label{fig:sc3_summary}
      \vspace*{-1.4em}
\end{figure}

\cref{fig:sc3_summary} shows similar measurement results as in the laboratory setting for the case when both antennas are directed to each other  in \cref{fig:sc3_messraum}. As the transmitter's antenna starts physically rotating and the beam alignment is progressively lost, the upper and lower DER bounds approach $0.5$, improving the secrecy performance. Further, when the eavesdropper is located perpendicular to the transmitter (Position D), the secrecy level notably increases, showing that an eavesdropper adjacent to or behind of the legitimate transmitter has lower chances of extracting information from its channel readings.

\section{Conclusion} \label{sec:conclusions}

We experimentally analyzed the effect of beamsteering  on the secrecy performance of an eavesdropper placed in 
various indoor environments. Our measurements showed that Eve needs to ensure beam alignment for weakening the secrecy performance in LOS scenarios. As  the beam misalignment increases, so does the secrecy performance considerably. In the NLOS scenario, Eve is able to exploit reflected signals due to the main and side lobes, which introduce a non-negligible information leakage for a relatively high SNR regime above \SI{5}{dB}. The use of reflective intelligent surfaces or a reduction in the secrecy rate should 
improve the secrecy performance here. We mitigated this effect either by reducing the transmit power or by steering the beam farther away. In summary, we demonstrated that PLS can be realized for realistic over-the-air scenarios using FR2 mmWave frequencies in 5G NR, while the reduced computation time for 
estimating upper and lower bounds on the DER 
allows for a real-time measurement of the secrecy levels that can be exploited by future 6G deployments.

\section*{Acknowledgments}
L. Torres-Figueroa and J. Voichtleitner would like to thank the support of Dr. Timo Noack during the measurement campaigns within the facilities of Rohde \& Schwarz GmbH, as well as to Dr. Thomas Nitsche for his valuable insight when decoding IQ samples collected using the R\&S®SMW200A.

\bibliographystyle{IEEEtran}
\bibliography{noauthorhyphen,IEEEabrv,literature}

\end{document}